\documentclass[prl,twocolumn,showpacs,preprintnumbers,amsmath,amssymb,floatfix]{revtex4}

\usepackage{subfigure,graphicx,epsfig,amsmath,amsfonts,amssymb,xcolor,slashed,ulem,multirow}

\usepackage{relsize}
\usepackage{slashed}
\usepackage{color}
\usepackage{ulem}
\usepackage{tabu}

\usepackage{amsmath}
\usepackage{amssymb}
\usepackage{amsthm}
\usepackage{mathrsfs}
\usepackage{graphicx}
\usepackage{epstopdf}
\usepackage{fancyhdr}
\usepackage{array}
\usepackage[all]{xy}
\usepackage{eufrak}
\usepackage{euscript}
\usepackage{enumerate}
\usepackage{slashed}
\usepackage{hyperref}
\usepackage{subfigure} 
\usepackage{epstopdf} 

\usepackage{hyperref}
\hypersetup{pdftex,colorlinks=true,linkcolor=blue,citecolor=blue,menucolor=black,urlcolor=blue,filecolor=blue}

\newcommand{\bB}{\bar{B}}
\newcommand{\qvp}{\vec{q}\,'}

\newcommand{\tfca}{T^{\rm FCA}}

\begin{document}
\title{Prediction of new states from $D^{(*)}B^{(*)}\bB^{(*)}$ three-body interactions}

\author{J.~M.~Dias}
\email{jorgivan.morais@ific.uv.es}
\affiliation{Departamento de F\'{\i}sica Te\'orica and IFIC, Centro Mixto Universidad de Valencia-CSIC, Institutos de Investigaci\'on de Paterna, Apartado
22085, 46071 Valencia, Spain}
\affiliation{Instituto de F\'{\i}sica, Universidade de S\~{a}o Paulo, C.P. 66318, 05389-970 S\~{a}o Paulo, SP, Brazil}

\author{L.~Roca}
\email{luisroca@um.es}
\affiliation{Departamento de F\'isica, Universidad de Murcia, E-30100 Murcia, Spain}

\author{S.~Sakai}
\email{shuntaro.sakai@ific.uv.es}
\affiliation{Departamento de F\'{\i}sica Te\'orica and IFIC, Centro Mixto Universidad de Valencia-CSIC, Institutos de Investigaci\'on de Paterna, Apartado
22085, 46071 Valencia, Spain}

\preprint{}

\date{\today}

\begin{abstract}
 We study three-body systems composed of $D^{(*)}$, $B^{(*)}$ and
 $\bB^{(*)}$ in order to look for possible bound states or resonances. In order to solve the three-body problem, we use the fixed center approach for the Faddeev equations considering 
 that the $B^*\bB^*(B\bB)$  are clusterized systems,  generated dynamically, 
 which interact with a third particle $D(D^*)$ whose mass is much smaller  than the two-body bound states forming the cluster.
 In the $DB^*\bB^*$, $D^*B^*\bB^*$, $DB\bB$ and $D^*B\bB$ systems with
 $I=1/2$, we found clear bound state peaks with binding energies typically a few
 tens MeV and more uncertain broad resonant states about ten MeV above the threshold
 with widths of a few tens MeV.
\end{abstract}

\pacs{14.40.Rt,12.40.Yx, 13.75.Lb}

\maketitle

\section{Introduction}

The heavy flavor sector (both open and hidden)  has gained  renewed 
attention in the last years by the  hadron physics community, 
in part spurred by the wide increase of experimental results
 (see Ref.~\cite{Chen:2016spr} for a recent review).
In the meson sector, specially interesting has been the proliferation of
states which cannot be easily accommodated as genuine $q\bar q$, like 
many $XYZ$-type resonances (see, {\it e.g.}, 
Refs.~\cite{olsenxyz,xliuxyz,Hosaka:2016pey,Chen:2016qju} for some reviews).
In the baryon sector remarkably sound was the discovery of the 
pentaquark $P_c(4450)^+$ by the LHCb collaboration \cite{expPc4450}.
Most of the non $q\bar q$ interpretations of many heavy flavor meson 
resonances lie within the picture of tetraquarks 
\cite{Wu:2017weo,Wu:2016gas,polotetra,Chen:2016qju} or meson-meson 
molecules \cite{Gamermann:2006nm,
Molina:2009ct,Ozpineci:2013qza,Dias:2014pva,Xiao:2013yca,Sun:2012sy,Sun:2011uh,Ohkoda:2012hv}.
Recently, several extensions to the heavy flavor sector  in three-body 
systems like $\rho B^* \bar{B}^*$ \cite{Bayar:2015zba}, $\rho D^* \bar D^*$  
\cite{Bayar:2015oea,Xiao:2012dw}, $DKK$  ($DK\bar{K}$) \cite{DKK} 
and $BDD$ ($BD\bar{D}$) \cite{Dias:2017miz} have been carried out 
with the prediction of several resonant states. The traditional way to deal 
with the three-body scattering amplitude has been to solve the Faddeev 
equations  \cite{Faddeev:1960su}. However, these equations are usually 
impossible to solve exactly and one has to resort to approximate methods. 
This is a feature well known by  the nuclear and hadron physics community 
where the Faddeev equations have been widely used to account for three-nucleon 
systems \cite{Alt,Epelbaum} or systems involving mesons and baryons 
\cite{nogami,Ikeda:2007nz,MartinezTorres:2007sr,Jido:2008kp}) or three-meson 
systems \cite{Mennessier:1972bi,MartinezTorres:2008gy,MartinezTorres:2009xb}.

The three-body problem can be drastically simplified when two of the 
particles form a bound cluster which is not much altered by the interaction 
with the third particle. In such a case one can resort to the so-called Fixed 
Center Approximation  (FCA) to the Faddeev equations 
\cite{Chand:1962ec,Barrett:1999cw,Deloff:1999gc,Kamalov:2000iy,Gal:2006cw}.
In the last years the FCA has proved its convenience in the study of many 
three-body systems in the light flavor sector 
\cite{Kamalov:2000iy,Roca:2010tf,YamagataSekihara:2010qk,Roca:2011br,Xie:2011uw,Bayar:2011qj}.
The first incursion in the charm sector with three-body resonances was 
done in Ref.~\cite{Xiao:2011rc} with the study of the $NDK$, $\bar{K} DN$ 
and $ND\bar{D}$ systems and also in Ref.~\cite{30} for $DNN$.

More recently, and involving only mesons, the FCA has been used to evaluate possible 
molecular states with open charm in $DKK$ and $DK\bar{K}$ \cite{DKK}, 
open bottom, open or hidden charm and double charmed three meson 
systems $BD\bar{D}$ and $BDD$ \cite{Dias:2017miz}. In the $DKK$ and 
$DK\bar{K}$ systems the evaluation using the FCA benefits from the fact 
that the $DK$ system is bound generating the $D_{s0}^*(2317)$ 
\cite{36,Gamermann:2006nm,38} and then the third particle rescatters 
with the components of the $DK$ cluster without breaking it. In the $BDD$ 
and $BD\bar{D}$ cases, the situation is analogous to the $DKK$ and 
$DK\bar{K}$ systems since the $BD$ system also bounds \cite{Sakai:2017avl}.

In the present work we analyze the $DB^*\bB^*$, $D^*B^*\bB^*$, 
$DB\bB$, and $D^*B\bB$ systems with $I=1/2$ to look for possible bound 
and/or resonant three-body states. In this case, the use of the FCA to 
evaluate the three-body scattering amplitude is suitable and appropriate 
since the $B\bar B$ and $B^*\bB^*$ systems in isospin $I=0$ were found to 
bound
\cite{Ozpineci:2013qza},
forming states of mass about $10450$ and 
$10550$~MeV, respectively. That corresponds to binding energies of about 100~MeV.
The work of
Ref.~\cite{Ozpineci:2013qza}
was based on the implementation of 
coupled channel unitary dynamics with kernels obtained from Lagrangians 
that combine local hidden gauge symmetry and heavy quark spin symmetry. 
In addition, for our present problem, we can also benefit from the fact that in 
Ref.~\cite{Sakai:2017avl} an attractive interaction, even producing bound 
states, was found for $BD$, $B^*D$, $BD^*$, $B^*D^*$, $B\bar D$, $B^*\bar D$, 
$B\bar D^*$ and $B^*\bar D^*$ in isospin $I=0$, with less binding energy than 
in the $B\bar B$ or $B^*\bB^*$ cases. On the contrary the analogous 
two-body interactions in isospin $I=1$ are repulsive, when allowed. However, 
since the $I=1$ amplitude is non-resonant one could expect a priori that 
the $I=0$ interaction will prevail, helping to bound the three-body state.


\section{Theoretical Framework}

In this section, we explain the formalism for the investigation of the
$D^{(*)}B^{(*)}\bB^{(*)}$ system. In the following, and in order to 
illustrate the process, we focus only on the $DB^*\bB^*$ case since 
we can obtain the expressions for the other channels in a similar way. 
As explained in the Introduction, in this study the FCA to the Faddeev 
equations is employed. This approach is effective when two of the three 
particles form a bound state, which will be called cluster, and there is 
not enough energy to excite the cluster \cite{MartinezTorres:2010ax}.
In the present calculation we are indeed in this situation since we are 
going to move in a range of energies close to the three-body (cluster 
+ third-particle) threshold and also the mass of the third particle, the 
projectile, is much smaller than the components of the cluster. In our 
case, the cluster is the $B^*\bB^*$ system, which according to the 
findings of Ref.~\cite{Ozpineci:2013qza} forms a bound state, with a 
binding energy of about $100$~MeV. The projectile is a $D$ meson, 
whose mass is much smaller than the components of the cluster. This 
$D$ meson undergoes multiple interactions with each component of the 
cluster. In this way, we need the  two-body $DB^*$ and $D\bB^*$ 
amplitudes (see Eq.~\eqref{eq_tBD} below) which enter as an input in 
the Faddeev equations. We obtain the $DB^*$ and $D\bB^*$ two-body 
amplitudes from Ref.~\cite{Sakai:2017avl}, based on a vector-meson 
exchange model from hidden gauge symmetry \cite{Bando,Meissner,Nagahiro} 
and implementing a unitarization procedure by means of the Bethe-Salpeter 
equation.

\begin{figure*}[t]
\centering
\includegraphics[width=15cm]{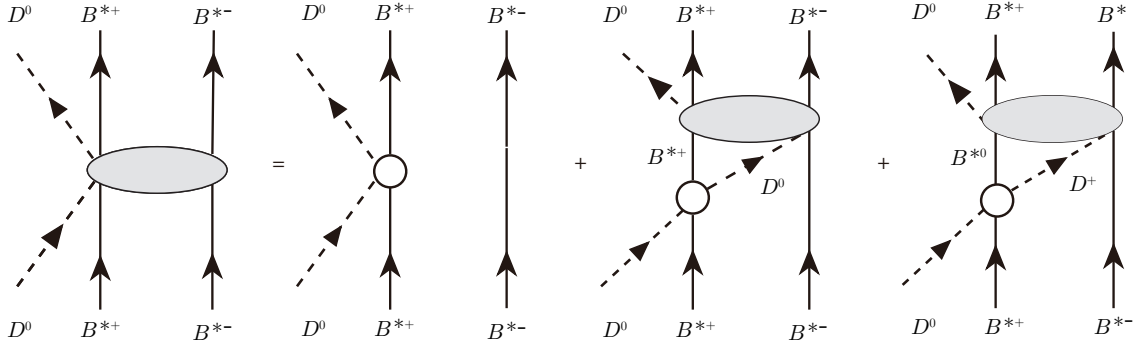}
\caption{Diagrams associated with the three-body amplitude for the 
$D^0B^{*+}B^{* -}$ interaction, contributing to the partition $T_{11}$ of the 
three-body amplitude.}
\label{diagrams}
\end{figure*}

In order to write the Faddeev equations with the FCA for the present case, 
we need to account for all the three-body diagrams contributing to the 
$DB^*\bB^*$ interaction. Since the scattering amplitude is independent 
of the third component of isospin, $I_3$, let us take, for example, the 
$I=1/2$, $I_3=-1/2$ case, for which we use the following nomenclature 
for the different channels needed:
\begin{align}
\label{channels}
 1)\ D^0[B^{*+}B^{*-}],& & 2)\ D^0[B^{*0}\bB^{*0}],
 & &3)\ D^+[B^{*0}B^{*-}],\nonumber\\
 4)\ [B^{*+}B^{*-}]D^0,& & 5)\ [B^{*0}\bB^{*0}]D^0,
 & & 6)\ [B^{*0}B^{*-}]D^+,
\end{align}
where the two particles in the brackets form the cluster
whose mass will be denoted by $M_c$, and the external $D$ meson 
is scattered first by the nearby particle, $e.g.$, the $D$ meson at the 
left-hand side of the bracket is scattered first by $B^*$, while the 
one at the right-hand side is scattered first by $\bB^*$.
Following this nomenclature, we can define the partition functions 
$T_{ij}$ which are the amplitudes for the diagrams accounting for the 
transition from the $i$  to the $j$ channels aforementioned, (see Eq.~\eqref{channels}). 
For instance, the amplitude associated with the transition of $D^0[B^{*+}B^{*-}]$ 
to itself, denoted by $T_{11}$, is given by the diagrams depicted in Fig.~\ref{diagrams}.
From this figure, we have
\begin{align}
T_{11}(s) = t_1(s_{DB^*}) +& t_1(s_{DB^*}) G_0(s) T_{41}(s) \nonumber \\
+& t_2(s_{DB^*}) G_0(s) T_{61}(s) \, , 
\label{T11}
\end{align}
where $s$ is the total three-body center-of-mass energy squared, while 
$t_1$ and $t_2$ are, respectively, the two-body $t_{D^0B^{*+}, 
D^0B^{*+}}$ and $t_{D^0B^{*+}, D^+B^{*0}}$ scattering amplitudes, in the 
charge basis, which can be easily related to the $DB^*$ amplitudes in isospin 
basis studied in Ref.~\cite{Sakai:2017avl}. These two-body amplitudes depend 
on the energy squared of the two-body subsystem, $s_{DB^*}$, 
(see Eq.~\eqref{eq_presc1} below). The $G_0$ function in the second 
and third terms of the right-hand side of Eq.~\eqref{T11} is the Green function 
of the $D$ meson between the particles of the cluster 
\cite{YamagataSekihara:2010qk}, given by
\begin{align}\label{G0}
 G_0(q^0)=\frac{1}{2M_c}\int\frac{d^3\vec{q}}{(2\pi)^3}\frac{F(\vec{q}\,)}
 {(q^{0})^2-\omega_D^2(\vec{q}\,)+i\epsilon} \,\, ,
\end{align}
with $\omega_D(\vec{q}\,)=\sqrt{|\vec{q}\,|^2+m_D^2}$. The energy carried 
by the $D$ meson between the components of the cluster, denoted by $q^0$, 
is a function of the total energy squared $s$, defined by
\begin{align}
 q^0=\frac{1}{2M_c}(s-m_D^2-M_c^2).
\end{align}

The information about the $B^*\bB^*$ bound state is encoded in the 
form factor $F(\vec{q}\,)$ appearing in Eq.~\eqref{G0}, which is related to the 
cluster wave function, $\Psi_c(\vec r\,)$, by means of a Fourier transformation, 
as it was discussed in Refs.~\cite{Roca:2010tf,YamagataSekihara:2010pj}:
\begin{align}
F(\vec q\,)=\int d^3\vec r \, e^{-i\vec q\cdot\vec r} \Psi_c^2(\vec r\,),
\end{align}
which can be obtained by
\begin{align}\label{eq_ff}
 F(\vec{q}\,)=\frac{1}{N}\int_V
 d^3\qvp\frac{1}{M_c-\omega_{B^*}(\qvp)-\omega_{\bB^*}(\qvp)}
 \nonumber \\ 
 \times\frac{1}
 {M_c-\omega_{B^*}(\vec{q}-\qvp)-\omega_{\bB^*}(\vec{q}-\qvp)}\, ,
\end{align}
where $V$ specifies the conditions $|\qvp|<\Lambda$ and $|\vec{q}-\qvp|<\Lambda$,
with $\Lambda$ the cutoff chosen to coincide with 
the value used in the evaluation of the $B^*\bB^*$ 
bound state \cite{Ozpineci:2013qza}. The normalization factor $N$ in Eq.~\eqref{eq_ff} is 
fixed such that $F(\vec{q}=0)=1$, and thus it is given by
\begin{eqnarray}\label{norm}
N = \int\limits_{|\vec{p}\,|<\Lambda}\, d^3 \vec{p}\, \Big( \frac{1}
{M_c  - \omega_{B^*}(\vec{p}\,) - \omega_{\bB^*}(\vec{p}\,)} \Big)^2 \, .
\end{eqnarray}
In Eqs.~\eqref{eq_ff} and \eqref{norm} we have 
$\omega_{B^*(\bB^*)}(\vec{p}\,)=\sqrt{|\vec{p}\,|^2+m^2_{B^*(\bB^*)}}$.

Following a similar procedure to the one used above to obtain the 
amplitude $T_{11}$ of Eq.~\eqref{T11}, we evaluate all the remaining 
amplitudes related to the transitions involving every channel listed in 
Eq.~\eqref{channels}, indicated by the indices $i,\,j$. Thus, we get a set 
of thirty-six coupled equations, since $i$ and $j$ run from $1$ to $6$, 
which provide the Faddeev equations with the FCA for the interaction we 
are concerned with. In matrix form, it reads
\begin{align}
 T=V+\widetilde{V}\,G_0\, T\label{eq_FCA_1} \, ,
\end{align}
where the matrices $V$ and $\widetilde{V}$ are
written in terms of the two-body $B^*D$ and $\bB^*D$ amplitudes as follows:
\begin{align}\label{VBDDbar}
 V =  
 \begin{pmatrix}
  t_{1} & 0 & t_{2} & 0 & 0 & 0 \\
  0 & t_{3} & 0 & 0 & 0 & 0 \\
  t_{2} & 0 & t_{4} & 0 & 0 & 0 \\
  0 & 0 & 0 & t_{5} & 0 & 0 \\
  0 & 0 & 0 & 0 & t_{6} & t_{7} \\
  0 & 0 & 0 & 0 & t_{7} & t_{8} 
 \end{pmatrix},\
 \widetilde{V} =
 \begin{pmatrix}
  0 & 0 & 0 & t_{1} & 0 & t_{2} \\
  0 & 0 & 0 & 0 & t_{3} & 0 \\
  0 & 0 & 0 & t_{2} & 0 & t_{4} \\
  t_{5} & 0 & 0 & 0 & 0 & 0 \\
  0 & t_{6} & t_{7} & 0 & 0 & 0 \\
  0 & t_{7} & t_{8} & 0 & 0 & 0 
 \end{pmatrix},
\end{align}
with
\begin{equation}\label{eq_tBD}
  \begin{array}{@{\,}ll@{\,}}
    t_{1}=t_{B^{*+}D^0,B^{*+}D^0}\,; & t_{5}=t_{B^{*-}D^0,B^{*-}D^0}\,;  \\
    t_{2}=t_{B^{*+}D^0,B^{*0}D^+}\,; & t_{6}=t_{\bB^{*0}D^0,\bB^{*0}D^0}\,;  \\
    t_{3}=t_{B^{*0}D^0,B^{*0}D^0}\,; & t_{7}=t_{\bB^{*0}D^0,B^{*-}D^+}\,;  \\
    t_{4}=t_{B^{*0}D^+,B^{*0}D^+}\,; & t_{8}=t_{B^{*-}D^+,B^{*-}D^+}\, .\\
  \end{array}
\end{equation}
These scattering matrix elements correspond to the two-body 
amplitudes for $DB^*$ and $D\bB^*$ interactions given in Ref.~\cite{Sakai:2017avl}.
In that reference the kernel of the unitarization procedure is obtained by 
the evaluation of mechanisms accounting for vector meson exchange from 
Lagrangians obtained from suitable extensions of hidden gauge symmetry 
Lagrangians to the heavy flavor sector, and compatible with the  heavy 
quark spin symmetry (HQSS) of QCD \cite{hqss}. The unitarization procedure only 
depends on one independent parameter, the three-momentum cutoff of 
the meson-meson loop function which turned out to be the largest source 
of uncertainty in Ref.~\cite{Sakai:2017avl}. We will also consider the 
uncertainty from that source in the results below. It is worth mentioning 
that the $I=0$ potential is attractive \cite{Sakai:2017avl} to the point to 
produce bound states   for $BD$, $B^*D$, $BD^*$, $B^*D^*$, 
$B\bar D$, $B^*\bar D$, $B\bar D^*$ and $B^*\bar D^*$. This is not the
case for $I=1$ where the lower order interaction is repulsive for
$\bar{B}^{(*)}D^{(*)}$ and zero for $B^{(*)}D^{(*)}$ \cite{Sakai:2017avl}.

The amplitudes $t_{1}$ -- $t_4$ and $t_5$ -- $t_{8}$ of Eq.~(\ref{eq_tBD}) 
must be multiplied by the normalization factors $c_1=M_c/m_{B^*}$ and 
$c_2=M_c/m_{\bB^*}$ respectively, to match the Mandl-Shaw normalization 
\cite{MandlShaw:2010} that we use. In this calculation, the polarization 
vectors of the vector mesons $B^*$ and $\bar B^*$ (or $D^*$ and $\bar D^*$ 
below) can be factored out in the two-body amplitudes since their consideration 
only gives a subleading contribution of the order of the squared momentum 
of the hadron over its mass \cite{Oset:2009vf,Roca:2005nm}.

The two-body amplitudes of Eq.~(\ref{eq_tBD}) depend on the energy of the 
corresponding two-body subsystem, $s_{ij}$, with $i$ the projectile and $j$ the 
corresponding particle of the cluster involved in the amplitude. In terms of the 
total three-body invariant mass squared, $s$, it is given by 
\cite{YamagataSekihara:2010qk,Bayar:2015zba}
\begin{align}
 s_{DB^*}=&m_D^2+m_{B^*}^2 \nonumber \\ +&\frac{1}{2M_c^2}\left(s-m_{D}^2-M_c^2\right)
 \left(M_c^2+m_{B^{*}}^2-m^2_{\bB^*}\right)\, . \label{eq_presc1}
\end{align}
The two-body energy of the $D\bB^*$ subsystem, $s_{D\bB^*}$, is 
obtained replacing the $B^*$ mass by the $\bB^*$ one in Eq.~(\ref{eq_presc1}). 
(Despite we have in this case $m_{B^{*}}=m_{\bB^*}$ (obviously), we keep 
them in Eq.~(\ref{eq_presc1}) just to know the general expression for other 
cases which could have different masses).

With all these ingredients, Eq.~(\ref{eq_FCA_1}) can be algebraically solved as
\begin{eqnarray}
T=(1-\widetilde{V}G_0)^{-1}V.\label{eq_FCA_res}
\end{eqnarray}

Finally, the three-body amplitude $T_{DB^*\bB^*}$ with $I=1/2$, associated 
with a $D$ meson interacting with the $B^*\bB^*$ ($I=0$) cluster, in terms 
of the matrix elements of $\tfca$ in Eq.~\eqref{eq_FCA_res} is
\begin{align}\label{Ttotal}
\nonumber T_{DB^*\bB^*}=&\frac{1}{2}\left(
T_{11}+T_{12}+T_{14}+T_{15}
+T_{21}+T_{22}+T_{24}+ T_{25}\right.\\
&\left.+T_{41}+T_{42}+ T_{44}+T_{45}
+T_{51}+T_{52}+T_{54}+T_{55}
\right).
\end{align}
This expression can be explicitly worked out in terms of the two-body 
amplitudes in isospin basis and gives
\begin{align}
&T_{DB^*\bar B^*}(s)=  \nonumber \\
&\ \frac{t_{\bar B^* D}^0+3t_{\bar B^* D}^1 
+(1-G_0t_{\bar B^* D}^0)(1+3G_0t_{\bar B^* D}^1)t_{B^* D}^0}
{4-G_0^2(t_{\bar B^* D}^0+3t_{\bar B^* D}^1) t_{B^* D}^0}
\label{eq:T3}
\end{align}
where $t_{\bar B^* D}^0\equiv t_{\bar B^* D, \bar B^* D}^{I=0}$,
$t_{\bar B^* D}^1\equiv t_{\bar B^* D, \bar B^* D}^{I=1}$ and 
$t_{ B^* D}^0\equiv t_{ B^* D,  B^* D}^{I=0}$.

For the other channels, $D^*B^*\bB^*$, $DB\bB$ and $D^*B\bB$, the 
procedure is analogous but changing the masses of the corresponding 
particles, and using the proper two-body amplitudes for the particles 
involved.

\section{Results}

For the numerical evaluation of the three-body amplitudes we use the 
following values for the meson masses: $m_{B}=5279.0$~MeV, 
$m_D=1869.0$~MeV, $m_{B^*}=5325.0$ MeV and $m_{D^*}=2007.0$ MeV.
As mentioned above, for the evaluation of the form factor of 
Eq.~\eqref{eq_ff}, which takes into account the clustering effect 
in the meson exchange between the constituents of the cluster, 
Eq.~\eqref{G0}, we  need the regularization cutoff $\Lambda$ which 
is conceptually analogous to the regularization cutoff used in the 
unitarization of the $[B\bB]$ and $[B^*\bB^*]$ in Ref.~\cite{Ozpineci:2013qza}. 
Since this cutoff is a free parameter of the model, one has to resort to 
some experimental result to constrain it. For instance, in 
Ref.~\cite{Gamermann:2006nm} a cutoff  of $415$~MeV was required to 
get a bound state at the experimental value of $3720$~MeV for the 
$D\bar{D}$ system. In Ref.~\cite{Nieves:2011vw} it was justified that 
heavy quark symmetry implies that the value of the cutoff is independent 
of the heavy flavor, up to corrections of order ${\cal O}(1/m_Q)$, with 
$m_Q$ the mass of the heavy quark. Therefore, in this line, in 
Ref.~\cite{Ozpineci:2013qza} a range of values between $415-830$~MeV 
for the $[B\bB]$ and $[B^*\bB^*]$ cutoff was justified, when compared 
to the cutoffs needed to obtain the $D\bar D$ resonance in 
Ref.~\cite{Gamermann:2006nm} and the the $D\bar D^*$ producing the 
$X(3872)$ in Ref.~\cite{Nieves:2011vw}. We will call this cutoff $\Lambda_{BB}$ 
in the following. Similar arguments were used in Ref.~\cite{Sakai:2017avl} to 
justify the use of a cutoff in the range $400-600$~MeV for the regularization 
of the $BD$-type interactions ($BD$, $B^*D$, $BD^*$, $B^*D^*$, 
$B\bar D$, $B^*\bar D$, $B\bar D^*$ and $B^*\bar D^*$). We will 
call this cutoff $\Lambda_{BD}$ in the following. Therefore, the variation of 
the cutoffs within the ranges $\Lambda_{BB}\sim 415-830$~MeV and 
$\Lambda_{DB}\sim 400-600$~MeV will be used to estimate the 
uncertainties in our approach. We also need the masses of the clusters 
$[B\bB]$ and $[B^*\bB^*]$, which are given by $M_{B\bB}=10523$~MeV and 
$M_{B^*\bB^*}=10613$~MeV for $\Lambda_{BB}=415$~MeV and 
$M_{B\bB}=10380$~MeV and $M_{B^*\bB^*}=10469$~MeV for 
$\Lambda_{BB}=830$~MeV \cite{Ozpineci:2013qza}.

\begin{figure}[t]
 \includegraphics[width=.95\linewidth]{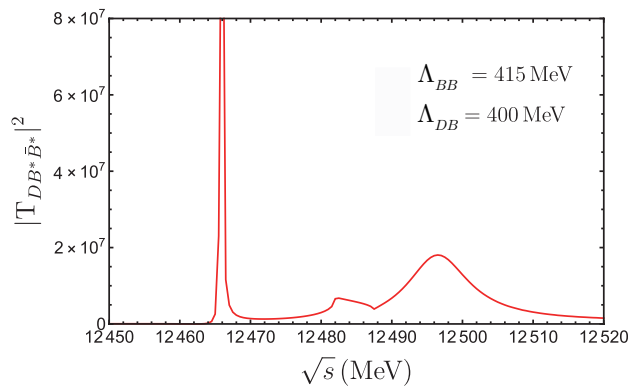}
 \caption{The three-body amplitude $|T_{DB^*\bB^*}|^2$ with
 $\Lambda_{DB}=400$~MeV and $\Lambda_{BB}=415$~MeV.}
 \label{fig_amp_tmp}
\end{figure}

As an example of the shape of the three-body amplitudes that we obtain, 
we show in Fig.~\ref{fig_amp_tmp} the squared amplitude of the 
$DB^*\bB^*$ three-body system, $|T_{DB^*\bB^*}|^2$, as a function 
of $\sqrt{s}$ for some particular values of the regularization cutoffs.
As we can see, we get a sharp peak at $\sqrt{s}=12466$~MeV,
which is
below the $D[B^*\bB^*]$ threshold, at $12482$~MeV.
This peak can be considered as a three-body 
$D[B^*\bB^*]$ bound state, with a binding energy of $16$~MeV.
Qualitatively similar plots, with different positions of the peaks, are 
obtained for the other three-body channels and different values of the 
regularization parameters.
\begin{table}[t]
 \centering
 \begin{tabular}[t]{c|c|c}
 \hline\hline
  &~$\sqrt{s_p}$ &$ E_B$ \\ \hline
  $DB^*\bB^*$~&~$12384\pm 65$&$27\pm 23$ \\
  $D^*B^*\bB^*$~&~$12520\pm 65$&$28\pm 25$ \\
  $DB\bB$~&~$12294\pm 64$&$27\pm 24$ \\
  $D^*B\bB$~&~$12430\pm 64$&$28\pm 25$\\ 
  \hline\hline
 \end{tabular}\\\vspace{3mm}
 \caption{Pole position $\sqrt{s_p}$ and
 the binding energy $E_B$ of ${D^{(*)}B^{(*)}\bB^{(*)}}$.
 The units are MeV.}
\label{table_pole}
\end{table}
This is summarized in Table~\ref{table_pole}, where we show
the positions of the poles below threshold for the different channels
obtained averaging over the results for the different values of the 
cutoffs within the ranges explained above. We also show in the last 
column the corresponding binding energies, $E_B$.
The emergence of these three-body bound states is quite robust
in our approach  since we obtain poles for all the values of the different 
cutoffs considered. Indeed the value for the upper limit of the $\Lambda_{BB}$ 
range ($830$~MeV) is a very conservative overestimation 
\cite{Ozpineci:2013qza} of this parameter and in spite of that we 
still get poles for that value of this cutoff.

It is important to note that the binding energies of these systems are
almost the same between different channels for the same set of 
regularization cutoffs. This is a non-trivial result and it is a consequence 
of the fact that the vector-meson exchange approach for the two-body 
interactions of $B^*D$, $\bB^*D$, and $B^*\bB^*$ respects the
HQSS.
Thus we can understand
this coincidence of the binding energy as a manifestation of the HQSS which has 
already seen in the two-body systems \cite{Ozpineci:2013qza,Sakai:2017avl,Lu:2014ina,Altenbuchinger:2013vwa}.

\begin{table}[t]
\begin{tabular}[t]{c|c|c}
\hline\hline
 &~$\Lambda_{DB}=400$~MeV ~&~$\Lambda_{DB}=600$~MeV  \\
 &$m_R\mid \Gamma$ &$m_R\mid \Gamma$ \\\hline
 $DB^*\bB^*$&$12497\mid 10$ &$12494\mid 15$ \\
 $D^*B^*\bB^*$&$12634\mid 15$ &$12632\mid 20$ \\
 $DB\bB$&$12407\mid 10$ &$12403\mid 15$ \\
 $D^*B\bB$&$12544\mid 15$ &$12542\mid 20$ \\
 \hline\hline
\end{tabular}
 \caption{Peak position and width of the resonances. $\Lambda_{BB}$ 
 is always fixed to $415$~MeV.}
 \label{tab_2}
\end{table}

On the other hand, in Fig.~\ref{fig_amp_tmp}, we find a broad bump 
located around $12500$~MeV and a width of the order of $10$~MeV 
in addition to the bound state previously discussed. Although this resonant 
state is above the $D[B^*\bB^*]$ threshold, it is still below the uncorrelated 
$DB^*\bB^*$ threshold. In Table~\ref{tab_2}, we summarize the energy 
of the peak position $m_R$ and the width $\Gamma$ for the $DB^*\bB^*$, 
$D^*B^*\bB^*$, $DB\bB$, and $D^*B\bB$ systems. The results on this table 
are obtained for the two extreme values of the cutoff $\Lambda_{BD}$ but 
only for one value of the cutoff $\Lambda_{BB}=415$~MeV. This is because 
we do not find a resonant structure above threshold for $\Lambda_{BB}=830$~MeV.
Therefore, the existence of the possible resonant state appearing above 
threshold in some specific cases are more uncertain than the bound
states
found below threshold, and further study would be necessary for clarification.

At this point it is worth clarifying some issues regarding the origin of the 
poles and resonances obtained above. First of all it is curious to note that for 
$D[B^*\bB^*]$
with $\Lambda_{BB}=415$~MeV and $\Lambda_{DB}=400$~MeV
the pole at $\sqrt{s}=12466$~MeV
that is shown in Fig.~\ref{fig_amp_tmp}
coincides exactly with the 
value of $\sqrt{s}$ for which  $\sqrt{s_{DB^*}}= 7175$~MeV, which is the pole 
position of $t_{B^* D}^0$. Usually, in other three-body problems, if there is a 
pole in the two-body amplitude, it does not manifest in the three-body 
amplitude since it cancels between the numerator and the denominator 
of the analogous expression to Eq.~\eqref{eq_FCA_res} or Eq.~\eqref{eq:T3}. 
However this is not the case for the present channels due to a subtle  accidental 
coincidence. Indeed, for $\sqrt{s}=12466$~MeV, $t_{B^* D}^0(s_{DB^*})$ has a 
pole and thus close to this energy Eq.~\eqref{eq:T3} reduces to
\begin{align}
T_{DB^*\bar B^*}(s)\simeq -\frac{(1-G_0t_{\bar B^* D}^0)(1+3G_0t_{\bar B^* D}^1)t_{B^* D}^0}
{G_0^2(t_{\bar B^* D}^0+3t_{\bar B^* D}^1)t_{B^* D}^0}
\label{eq:T3reduced}
\end{align}
Note that $t_{B^* D}^0$ cancels between numerator and denominator 
and therefore Eq.~\eqref{eq:T3} should have no pole. However, it turns out 
that, by coincidence, $t_{\bar B^* D}^0+3t_{\bar B^* D}^1$ has a zero at exactly 
the same value where $t_{B^* D}^0$ has the pole. In order to see why this 
happens, let us note that the  $B^*D$ potentials have the structure 
\begin{align}
V_{BD}^I= \alpha \,a
\end{align}
where $a=-\frac{1}{8 f^2}\left[ 
      3s_{DB^*}-2(m_{B^*}^2+m_{D}^2)-\frac{(m_{B^*}^2-m_{ D}^2)^2}{s_{DB^*}}\right]$ 
      and $\alpha$ is a coefficient which value is  $\alpha=2$ for  $V_{B^* D}^0$, $\alpha=1$ for 
$V_{\bar B^* D}^0$ and $\alpha=-1$ for $V_{\bar B^* D}^1$ (see Eqs.~(15)-(17) in Ref.~\cite{Sakai:2017avl}).
Therefore, the two-body unitarized amplitudes are given by \cite{Sakai:2017avl}
\begin{align}
t_{B^* D}^0=\frac{2a}{1-2aG}
\label{eq:5}
\end{align}
with $G$ the $B^*D$ loop function.
Equation~\eqref{eq:5} has a pole when
\begin{align}
1-2aG=0.
\end{align}
But, on the other hand we have
\begin{align}
t_{\bar B^* D}^0=\frac{a}{1-aG},
\label{eq:kk1}
\end{align}
\begin{align}
t_{\bar B^* D}^1=\frac{-a}{1+aG}
\label{eq:kk2}
\end{align}
and therefore 
\begin{align}
t_{\bar B^* D}^0+3t_{\bar B^* D}^1= \frac{a}{1-aG}-\frac{3a}{1+aG}
=\frac{-2a(1-2aG)}{1-(aG)^2}
\end{align}
which is zero when $1-2aG=0$ which is, by accident, exactly the same condition for 
$t_{B^* D}^0$ to have a pole, (see Eq.~\eqref{eq:5}). It is worth noting that, if 
the model for the two-body amplitudes were a bit different, {\it e.g.} considering 
subleading terms in $t_{\bar B^* D}^1$, for instance, then the three-body pole would 
not coincide exactly with the two-body pole of $t_{B^* D}^0$. We have checked that 
even changing $t_{\bar B^* D}^1$ by hand about 20\%, the three-body pole still 
appears but at an slightly different position. Therefore this pole has to be considered 
as an actual  three-body state since it corresponds to a pole of Eq.~\eqref{eq:T3}, where 
the two-body pole cancels. Thus the pole in the three-body amplitude has nothing to do 
with the two-body pole even though it coincides numerically in the position for the 
channels considered in the present work. On the other hand, we are going to justify 
that the bump above threshold comes also from the thee-body dynamics and is 
related to a different pole of Eq.~\eqref{eq:T3}. Indeed, the possible poles of Eq.~\eqref{eq:T3} 
would correspond to zeroes of its denominator:
\begin{align}
4-G_0^2(t_{\bar B^* D}^0+3t_{\bar B^* D}^1) t_{B^* D}^0=0.
\label{eq:T3denom}
\end{align}
Using Eqs.~\eqref{eq:5}, \eqref{eq:kk1} and \eqref{eq:kk2}, one obtains that 
Eq.~\eqref{eq:T3denom} has two solutions, one when 
\begin{align}
1-2aG=0,
\end{align}
which is the solution that produces the pole below threshold,
and the other solution when
 \begin{align}
a^2(G^2-G_0^2)-1=0,
\label{eq:condition2}
\end{align}
which produces the resonance above threshold. Actually we find that the
poles associated with Eq.~\eqref{eq:condition2} happen for complex $\sqrt{s}$ 
since they occur for  ${\rm Re}[\sqrt{s}\,]$ above the cluster + third-particle 
threshold. For the channels we are considering in the present work, we have 
checked that the  ${\rm Re}[\sqrt{s}\,]$ of the solution of Eq.~\eqref{eq:condition2} 
are close to the position of the maximum of the bump found in the three-body 
amplitudes. Therefore, and in summary, the bumps found above threshold 
should also be considered as three-body resonances since they correspond 
to poles of the three-body amplitude.

\section{Summary}

We have investigated theoretically the three-body interactions $DB^*\bB^*$, 
$DB^*\bB^*$, $DB\bB$ and $D^*B\bB$  taking into account dynamical models 
for the $D^{(*)}B^{(*)}$, $D^{(*)}\bB^{(*)}$  and $B^*\bB^*(B\bB)$ subsystems 
studied in previous works. This has allowed us to apply the fixed center approximation 
to the Faddeev equations where the $B^*\bB^*(B\bB)$ two-body subsystems are 
bound forming clusters, which then interact with a $D^{(*)}$ meson. As a result, 
we have found three-body bound states for each one of these systems with 
binding energies around $20-30$~MeV. This similarity in the binding between 
the different channels is a clear manifestation of the heavy quark spin symmetry. 
Furthermore, we have also found resonant bumps  above the $D^{(*)}[B^{(*)}\bB^{(*)}]$ 
threshold with width about $10$~MeV, however these bumps are not stable under 
the uncertainties that come from the cutoff values used to regularize the two-body 
meson-meson loops and then their existence are not so clear than the bound 
states below threshold.

\section*{Acknowledgments}
J.~M.~Dias would like to thank the Brazilian funding agency FAPESP for
the financial support under Grant No.~2016/22561-2.
S.~S. acknowledges the support of the Generalitat Valenciana in the program
Prometeo II-2014/068.



\end{document}